\documentclass[a4paper]{jpconf}
\usepackage{graphicx}
\usepackage{amsmath,amsfonts}
\usepackage{amssymb}
\usepackage{amscd}
\usepackage[font=small,labelfont=bf]{caption}
\usepackage{wrapfig}
\usepackage{color}
\makeatletter

\def\@makecaption#1#2{%
  \vskip\abovecaptionskip
  \hb@xt@\hsize{\hfil#2\hfil}%
  \vskip\belowcaptionskip}
\makeatother

\newcommand{\nn}{\nonumber}
\newcommand{\beq}{\begin{equation}}
\newcommand{\eeq}{\end{equation}}
\newcommand{\beqa}{\begin{eqnarray}}
\newcommand{\eeqa}{\end{eqnarray}}

\newtheorem{theorem}{Theorem}[section]

\newtheorem{definition}[theorem]{Definition}
\newtheorem{example}[theorem]{Example}

\newtheorem{proposition}[theorem]{Proposition}

\newtheorem{method}[theorem]{Method}
\begin{document}
\title{Dihedral Invariant Polynomials in the effective Lagrangian of QED}

\author{Idrish Huet}
\address{Facultad de Ciencias en F\'isica y Matem\'aticas, Universidad Aut\'onoma de Chiapas\\
 Ciudad Universitaria, Tuxtla Guti\'errez 29050, Mexico \\
 Theoretisch-Physikalisches Institut, Friedrich-Schiller-Universit\"at Jena,\\
Max-Wien-Platz 1, D-07743 Jena, Germany
 }
\author{Michel Rausch de Traubenberg}
\address{Universit\'e de Strasbourg, \\CNRS, IPHC UMR7178, \\F-67037 Strasbourg Cedex, France 
}
\author{Christian Schubert}
\address{Instituto de F{{\'\i}}sica y Matem\'aticas, Universidad Michoacana de San Nicol\'as de Hidalgo\\
Apdo. Postal 2-82, C.P. 58040, Morelia, Michoacan, Mexico\\
Kavli Institute for Theoretical Physics,
University of California, 
Santa Barbara, CA  93106, USA.}


\ead{idrish.huet@gmail.com, Michel.Rausch@iphc.cnrs.fr, Christianschubert137@google.com}

\begin{abstract}
We present a new group-theoretical technique to calculate weak field expansions for some Feynman diagrams using invariant polynomials of the dihedral group. In particular we show results obtained for the first coefficients of the three loop effective lagrangian of 1+1 QED in an external constant field, where the dihedral symmetry appears. Our results suggest that a closed form involving rational numbers and the Riemann zeta function might exist for these coefficients.
\end{abstract}

\section{Introduction}

The famous Euler-Heisenberg lagrangian (EHL) is the one-loop quantum correction to the Maxwellian lagrangian for classical electrodynamics in a constant field, it was elegantly cast as a proper-time integral by Euler and Heisenberg \cite{EH} already in 1936:

\[
\mathcal{L}^{(1)} = - \frac{1}{8\pi^2} \int_0^{\infty} \frac{dT}{T^3} e^{-m^2 T} \left[\frac{(eaT)(ebT)}{\tanh(eaT)\tan(ebT)} - \frac{1}{3} (a^2 - b^2) T^2 -1 \right].
\]
Written in terms of the invariants $ab = {\bf E \cdot B}, \quad a^2 - b^2 = B^2 - E^2$ and the fermion mass $m$, this representation already includes renormalisation of vacuum energy and charge. The EHL encodes information about remarkable quantum effects of an external constant electromagnetic field on the vacuum such as light-light scattering, field-dependence of the speed of light, Schwinger pair creation and vacuum birefringence among others \cite{Dunne,Dittrich}. Concerning light-light scattering the EHL provides also the low energy limit of the one-loop $N$-photon amplitudes \cite{MSV}, this becomes apparent from the diagrammatic expansion of the EHL:

\begin{figure}[h]
\begin{center}
\includegraphics[width=0.48\textwidth]{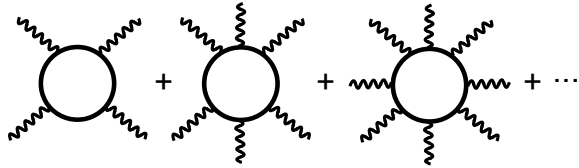}
\caption{Fig. 1. Diagrammatic expansion of the one-loop EHL}
\end{center}
\end{figure}
\noindent
in the diagrams above the external photons have all energies $\hbar \omega << m c^2$, since a constant field may only create zero energy photons. The main concern of this work is to present a new group theoretical approach to calculate multiloop corrections in the weak field limit to the EHL, in particular we demonstrate how this is done at the three-loop level.

The motivation behind this calculation is to investigate the asymptotic behaviour of the multi-loop EHL in the weak-field regime to probe a non-perturbative conjecture that presently stands based on previous results obtained by Borel summation methods \cite{DS} and worldline instanton techniques \cite{AAM}. The conjecture in question is an asymptotic relation between the imaginary part of the all-order EHL and the one-loop EHL in a weak purely electric field 

\begin{equation}
    \sum_{\ell=0}^{\infty} \mbox{Im} \mathcal{L}^{(\ell)} (E)~ \sim^{\!\!\!\!\!\!\!\! E \to 0} \mbox{Im}  \mathcal{L}^{(1)} (E) e^{\alpha \pi}
\end{equation}
Where $\mathcal{L}^{(\ell)} (E)$ is the $\ell$-loop correction to the EHL. This prediction, which we shall call the {\it exponential conjecture} is grounded on the Affleck-Alvarez-Manton formula \cite{AAM} and closely linked to P. Cvitanovic's conjecture on the convergence of the quenched series of QED \cite{Cvitanovic}.

For the case at hand we chose to investigate this prediction for the EHL of a constant field in the toy model 1+1 QED, this choice was motivated by the result of Krasnansky \cite{Kras}, who showed that even at $\ell =2$ the EHL for a constant field in 1+1 scalar QED has essentially the same structure as that of 3+1 scalar QED, and the realisation that self-dual fields provide substantial computational simplifications \cite{DunneSchubert, DS2,DS3}, our goal being to compare a self-dual field in 3+1 QED against a constant field in 1+1 QED.

The exponential conjecture can be extended to dimension $D=2$, in scalar QED this was done in \cite{HMcS,HMcS2}

\begin{equation} \label{exp2D}
    \sum_{\ell =0}^{\infty} \mbox{Im} \mathcal{L}^{(\ell)} (E)~ \sim^{\!\!\!\!\!\!\!\! E \to 0} \frac{eE}{4\pi} e^{-\frac{m^2 \pi}{eE} + \tilde{\alpha} \pi^2 \kappa^2}
\end{equation}
where $\tilde{\alpha} = \frac{2e^2}{\pi m^2}$ is the fine structure constant in two dimensions and $\kappa = m^2/2eE$. In the weak field regime the $\ell$-loop EHL for $D=2$ has the Taylor series expansion

\begin{equation}
\mathcal{L}_{2D}^{(\ell)} = \frac{m^2}{2\pi} \sum_{n=1}^{\infty} (-1)^{\ell-1} c^{(\ell)}_{2D}(n) (i\kappa)^{-2n}
\end{equation}
Contrary to $D=4$ QED in $D=2$ the asymptotic growth of the coefficients $c^{(\ell)}_{2D}(n)$ increases with $\ell$, this may be related to the fact that in two dimensions the Coulomb potential is confining. Borel analysis of (\ref{exp2D}) leads to the following condition on the coefficients

\begin{equation}
    \lim_{n \to \infty} \frac{c^{(\ell)}_{2D}(n) }{ c^{(1)}_{2D} (n+\ell-1)} = \frac{(\tilde{\alpha} \pi^2)^{\ell-1}}{(\ell-1)!}
\end{equation}
This prediction was verified in \cite{HMcS,HMcS2} at the level $\ell=2$. Further differences and analogies between $D=4$ and the technically far simpler $D=2$ toy model suggest pushing this calculation to the level $\ell=3$ and higher orders to learn more about the validity of the exponential conjecture. This question will be pursued fully in the upcoming work \cite{HRS}. An adequate approach to carry out the computation of the expansion coefficients is thus needed, and while different approaches might work we have proposed a technique based on the discrete symmetries of each Feynman graph. For definiteness consider the EHL $\mathcal{L}^{(3)}_{2D}$, it has dominant contributions from the Feynman diagrams A and B

\begin{figure}[h]
    \centering
    \includegraphics[width=0.45\textwidth]{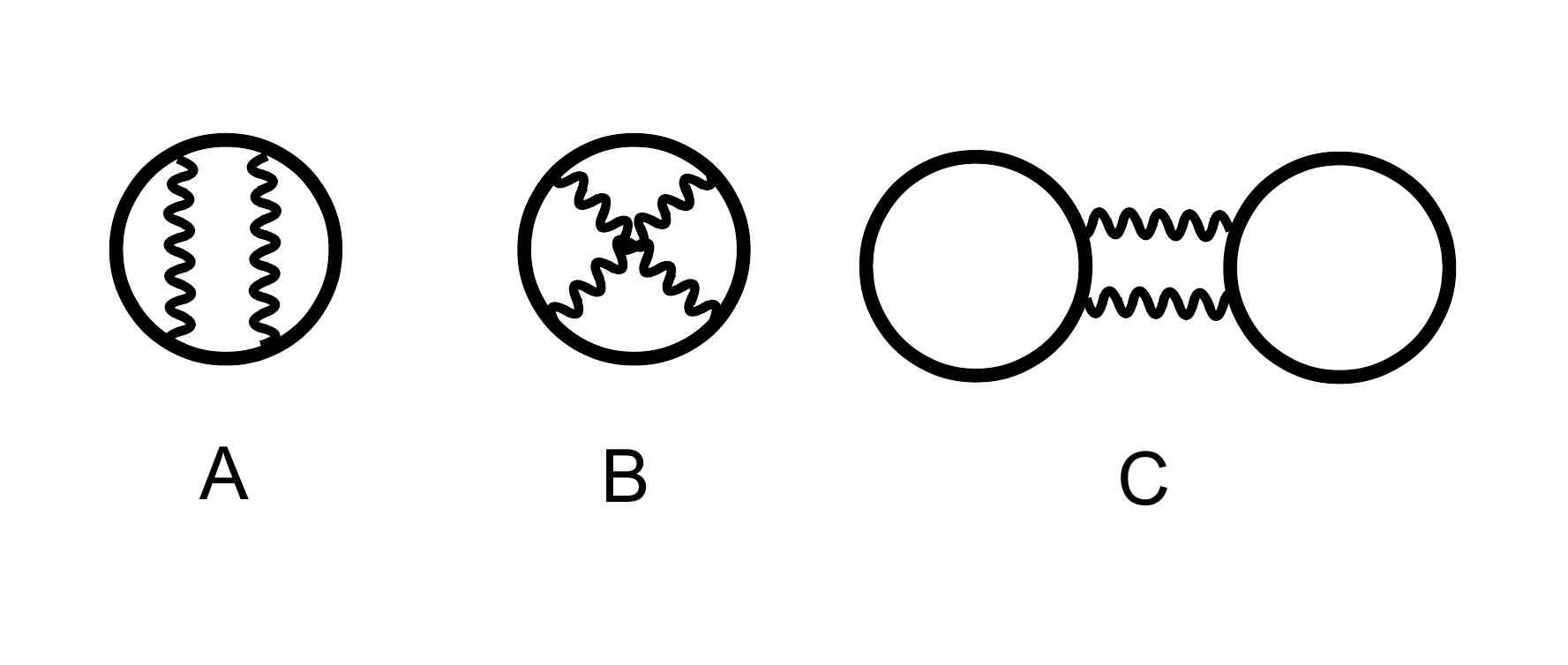}
    \caption{Fig. 2a. Dominant amplitudes in the three-loop EHL}
\end{figure}
and subdominant contributions coming from diagram C and the diagrams

\begin{figure}[h]
    \centering
    \includegraphics[width=0.45\textwidth]{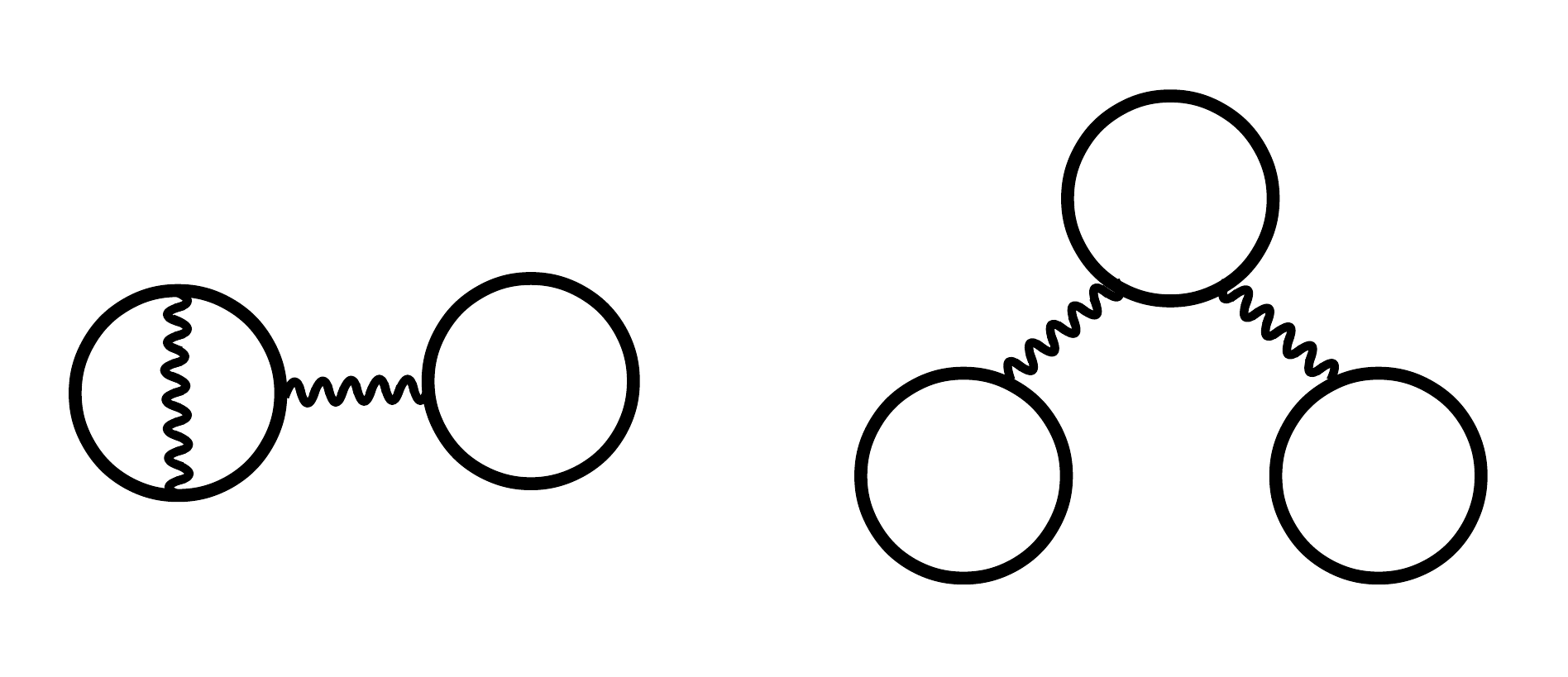}
    \caption{Fig. 2b. Subdominant amplitudes in the three-loop EHL}
\end{figure}
In these five diagrams the fermion lines represent the full propagator under the external field. Remarkably, and contrary to a long held belief, the tadpole diagrams in Fig. 2b are non-vanishing, as was first pointed out by an actual calculation done by H. Gies and F. Karbstein \cite{GK}. To exemplify the technique we shall center our efforts in calculating diagram B below, to this end we introduce the Schwinger parameters $z,z',\bar{z},\hat{z}$:
\begin{figure}[h]
    \centering
    \includegraphics[width=0.15\textwidth]{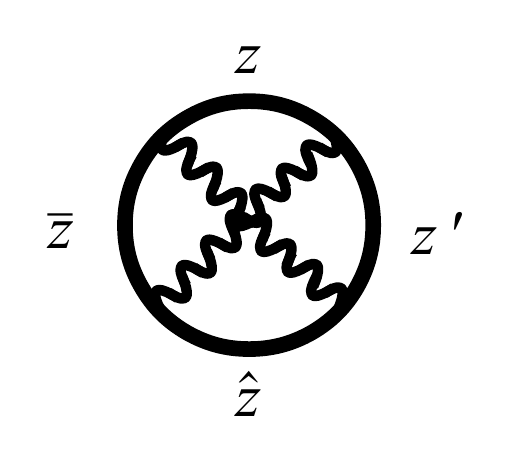}
    \caption{Fig. 3. Diagram B with dihedral symmetry $D_4$}
\end{figure}
the amplitude in diagram $B$ is then written as a fourfold integral over the Schwinger parameters

\begin{eqnarray}
 \Gamma^{(B)}=&& \frac{1}{32\pi^3}\frac{e^3}{f}\int_0^{\infty}\!\!\!  \int_0^{\infty}\!\!\!  \int_0^{\infty} \!\!\! \int_0^{\infty}dzdz'd\hat zd\bar z \,e^{-2\kappa (z+z'+\hat z+\bar z)}
 \nonumber \\ 
&& \biggl\lbrace  \frac{1}{\cosh^2 z \cosh^2 z' \cosh^2\hat z\cosh^2 \bar z}
\frac{B}{A^3C}  \nonumber \\ \nonumber &&
 - 4\kappa^2 \frac{\cosh({{ z- z' + \hat z - \bar z}})}{\cosh z \cosh z' \cosh \hat z\cosh \bar z}
\Bigl\lbrack
\frac{1}{A} - \frac{C}{G^2}\ln \Bigl(1+ \frac{G^2}{AC}\Bigr)
\Bigr\rbrack
\biggr\rbrace
\end{eqnarray}
where we used the following shorthands
\begin{eqnarray}
A &=& \tanh z + \tanh z' + \tanh \hat z + \tanh \bar z ,\nonumber \nn \\
B &=& (\tanh^2 z + \tanh^2 \hat z)(\tanh z' + \tanh \bar z) +   \nonumber \\ && (\tanh^2 z' + \tanh^2 \bar z)(\tanh z + \tanh \hat z), \nn \\
C &=& \tanh z\tanh z' \tanh \hat z + \tanh z \tanh z' \tanh \bar z +  \nonumber \\ && \tanh z \tanh \hat z \tanh \bar z + \tanh z' \tanh \hat z \tanh \bar z ,\nn \\
G &=& \tanh z \tanh \hat z - \tanh z' \tanh \bar z \nonumber 
\end{eqnarray}
a direct attempt to carry out the integrations, even after Taylor expansion has been carried out, is hopeless due to spurious singularities in the logarithm that symbolic computation software as MATHEMATICA fails to handle. This problem calls for another approach able to manage these singularities, and one such approach will be outlined in what follows.

\section{Group Theory in QFT}
As we have seen Diagram B is difficult to compute. The main idea to obtain the weak
field limit of this  diagram is to use its high degree of symmetry and to organise the expansion by
mean of group theory. In fact it can directly observed that diagram $B$ is invariant under the following transformations ($z=\rho w, z'=\rho w', \hat z = \rho \hat w, \bar z = \rho \bar w, \rho=\frac{ef}{m^2}$)  :
\beqa
\begin{array}{lclc}
g_1:&w &\leftrightarrow& \hat w \ , \\
g_2:&w' &\leftrightarrow&  \bar w \ , \\
g_3:&(w,\hat w) &\leftrightarrow& (w',\bar w)\ .
\end{array}
\eeqa
The group generated by the three transformations above is eight-dimensional, non-abelian and corresponds to the Dihedral group $D_4$.
In fact, the procedure we will present below is quite general and could be applied for Feynman diagrams presenting a high degree of symmetry in any QFT.

\subsection{Polynomial invariants for finite group}
In this subsection we briefly present the method.
Consider now $G$ a finite group and let $\rho(G)=\Gamma \subset GL(n,\mathbb  R)$ be an $n-$dimensional real representation of
$G$. The representation $\Gamma$ can be reducible or irreducible. Note also that $\Gamma$ is assumed to be
a real representation, but the method is equally valid for complex or pseudo-real representations.
Noting  $\mathbb R^n$ the carrier space of the representation $\Gamma$, for any  $X =(x_1,\cdots, x_n)^t\in \mathbb R^n$
the representation $\rho$  induces naturally an action onto the set of polynomials $\mathbb R[x_1,\cdots,x_n]$ with $n$ variables as
follows
\beqa
g. P(X) \equiv P(g.X) \nn \ ,
\eeqa
for any $g \in \Gamma$ with $g.X$ the natural action of $g$ on $X =(x_1,\cdots,x_n)^t \in \mathbb R^n$. 
\begin{definition}
  Let $P \in \mathbb R[x_1,\cdots,x_n]$, the polynomial $P$ is said to be invariant iff
  \beqa
\forall g \in \Gamma\ \ g\cdot P(X) = P(X) \ . \nn
\eeqa
The set of invariant polynomial is denoted
\beqa
 \mathbb R[x_1,\cdots,x_n]^\Gamma = \Big\{ I \in \mathbb R[x_1,\cdots,x_n] \ \  \text{s.t.} \ \ g\cdot I(X) = I(X) \ \ \forall g \in \Gamma \Big\} \ . \nn
\eeqa
\end{definition}

If we perform a Taylor expansion around zero of Molien's generating function  $\Phi(t)$
\beqa
\Phi(t) = \frac 1 {|G|} \sum \limits_{g \in \Gamma} \frac 1 {\det(1 -tg)} = \sum \limits_{k=0}^{+\infty} n_k t^k \ , \nn
\eeqa
the coefficients $n_k$ precisely give the number of linearly invariant polynomial of degree $k$. Of course it is obvious that these
polynomials are not algebraically independent. For instance any degree  $k$  invariant polynomial $I_k$ gives rise to a the degree $2k$ invariant
polynomial $I_k^2$. So one may naturally wonder if one can define a ``basis'' (in a sense to be defined) of polynomial invariants.
In fact we now recall that any invariant can be expressed in terms of the so-called primitive and secondary invariants
\cite{sturm-book}.

\begin{proposition}\label{prop:PS}
  Let $G$ be a finite group and let $\Gamma$ be an $n-$dimensional (real) representation.
  \begin{enumerate}
    \item There exists
  $n= \dim (\Gamma)$ algebraically invariants polynomals,
  $P_1,\cdots, P_n$, {\it i.e.}, such that the Jacobian
  \beqa
\frac{\partial(P_1,\cdots,P_n)}{\partial(x_1,\cdots,x_n)} \ne 0 \nn
\eeqa
called the primitive invariants. 
Denote $d_k=\deg(P_k)$ and ${\cal R} = \mathbb R[P_1,\cdots,P_n]$
the
subalgebra of  polynomial invariants generated by the primitive invariants.
\item There exists $m= d_1 \cdots d_n/|G|$
  secondary invariants polynomials $S_1,\cdots, S_m$.
\item The set of polynomials $(P_1,\cdots,P_n,S_1,\cdots, S_m)$ is not algebraically independant
  and satisfy algebraic relations called syzygies.
  \item The subalgebra of invariants $\mathbb R[x_1,\cdots,x_n]^\Gamma$ is a free ${\cal R}-$module with basis $(S_1,\cdots,S_m)$.
In particular this means that any invariant $I \in  \mathbb R[x_1,\cdots,x_n]^\Gamma$ can be uniquely written as
\beqa
\label{eq:PS}
I=\sum\limits_{i=1}^m  f_i(P_1,\cdots,P_n) S_i \ , 
\eeqa
where $f_i(P_1,\cdots,P_n), i=1,\cdots, m$ belongs to ${\cal R}$, {\it i.e.}, are polynomials in $(P_1,\cdots,P_n)$.
\end{enumerate}
\end{proposition}

The syzygies have a natural interpretation in terms of \eqref{eq:PS}. Consider for instance the secondary invariant polynomials
$S_i$, thus $S_i^2$ is an invariant polynomial, and by Proposition \ref{prop:PS}  there exists $f_{i,1},\cdots,f_{i,n} \in {\cal R}$
such that 
\beqa
S_i^2-\sum\limits_{j=1}^m  f_{i,j}(P_1,\cdots,P_n) S_j =0\ , \nn 
\eeqa
these relations among the  $P$s and  the $S$s are syzygies.

In general given a finite group $G$ and a representation $\Gamma$, it is not an obvious task to identify the primary and the
secondary polynomials. In fact there exist many automated ways to solve this problem. In this paper, we have
used  the Computer Algebra System for Polynomial Computations  called
SINGULAR \cite{decker}.

\begin{example}\label{ex:S}
  Let $\Sigma_4$ be the permutation group  acting on four elements and let $\Gamma=\mathbb R^4$ be the four-dimensional representation.
 The set of symmetric polynomials of degree $d=1,\cdots,4$ is the
set of primitive invariants and the only secondary invariant is $\sigma_0=1$. 
Introducing the symmetric polynomials 
\beqa
\sigma_1 &=& w + \hat w + w' + \bar w \ , \nonumber \\
\sigma_2&=& w \hat w + w w' + w \bar w +\hat w  w' + \hat w \bar w + w' \bar w\ , \nonumber \\
\sigma_3&=& w  \hat w  w'+ w  \hat  w  \bar w +w  w'  \bar w + w' \hat w \bar w\nonumber \\
\sigma_4&=& w \hat w w' \bar w \ . \nonumber\nn
\eeqa
we have for any invariant polynomial
\beqa
I(w,w',\bar w,\hat w)=P(\sigma_1,\sigma_2,\sigma_3,\sigma_4) \ . \nn
\eeqa
\end{example}

\begin{example} \label{ex:D}
  Let $D_4$ be the Dihedral group and let $\Gamma = \mathbb R^4$. It is obvious that in this case not all the $\sigma$s of Example
  \ref{ex:S} are invariant polynomials. In this case the Molien series takes the form:
 \beqa
    \Phi(t)&=& \frac 18 \Big(\frac 2{1-t^4} + \frac2 {(1-t)^3(1+t)}+\frac 3 {(1-t^2)^2} + \frac 1{(1-t)^4}\Big) \nn\\
    &=&1+t+3t^2+4t^3+8t^4+10t^5+16 t^6 +O(t^7)\nn
    \eeqa
 and there are {\it e.g.} four degree three invariant polynomials. 
Now,   using SINGULAR, we obtain the primitive invariants
\beqa
a&=&\sigma_1=  w + \hat w + w' + \bar w \ , \nonumber \\
\lambda&=&(w + \hat w)(w' + \bar w) \ ,\nonumber \\
\mu &=&w \hat w + w' \bar w  , \nonumber \\
d&=& \sigma_4=w \hat w w' \bar w \ , \nonumber
\eeqa
of degree $d_1=1, d_2= d_3=2,d_4=4$ respectively.
A direct computation gives
\beqa
\frac{\partial(a,\lambda,\mu,d)}{\partial(w,w',\hat w,\bar w)}=(w - w')(w - \bar w )(w -\hat w)(w'- \bar w )
(w'- \hat w )(\bar w - \hat w) \ . \nn
\eeqa
Thus the number of secondary invariants is
\beqa
m=\frac{1\times 2^2\times 4}{8} =2 \ . \nn
\eeqa
Using  SINGULAR,
again the secondary invariant
are given by
\beqa
\sigma_0=1 \ , \quad
c=\sigma_3= w w' \hat w + w w'\bar w + w \bar w \hat w + w' \hat w \bar w  \ .\nn
\eeqa
In this case, the syzygy is
 \beqa
\lambda(\mu^2-4 d)+ a^2d- a\mu c\ + c^2\ =0\ .\nn
    \eeqa
Thus any invariant polynomial uniquely writes like
\beqa
I(w,w',\bar w,\hat w)=f(a,\lambda,\mu,d) + g(a,\lambda,\mu,d) c \ , \nn
\eeqa
with $f,g \in \mathbb R[a,\lambda,\mu,d]$.
\end{example}

We have considered so far the so-called invariant polynomials. It turns out that there exists another class of polynomial:
\begin{definition}
  Let $P \in \mathbb R[x_1,\cdots,x_n]$, the polynomial $P$ is said to be semi-invariant iff
  \beqa
\forall g \in G, \ \ g.P(X)=e^{i\varphi(g)} \;P(X) \ ,\nn 
\eeqa
where the phase $\varphi(g)$ depends on the element $g$. Since we are concerned with real representations we have here
$e^{i\phi(g)} =\pm 1$ and $P$ is invariant up to a sign.
\end{definition}
The consideration of semi-invariant polynomials will be in turn essential to improve our algorithm.

\begin{method}\label{met:PI}
Considering a given Feynman diagram depending on some parameters and invariant under a finite group of symmetry, then the computation
of this Feynman diagram can be organised by means of the invariant polynomials (in the parameters on which the Feynman diagram depends) of $G$.
\end{method}
This is precisely Method \ref{met:PI} that  we will use to compute  diagram $B$ with some improvement involving semi-invariant polynomials.

\subsection{Application to diagramm $B$}
Method \ref{met:PI} is general, however in our particular case the method can be improved by the inclusion of semi-invariants. Thus, we now
    introduce
   \beqa
   \tilde w&=& w+ \hat w  - w' -\bar w\nn\\
  g&=& w \hat w - \bar w w' \nn\\
 j &=&( w -  \hat w - w' + \bar w)( w -  \hat w +  w' - \bar w)\nn\\
  j'&=& ( w-\hat w)(\bar  w-\ w') \ ,  \nn
  \eeqa
  which are semi-invariant. Indeed, 
  it can be checked easily that $\tilde w,g$ and $j$ change sign only under the action of $g_3$ whereas $j'$ changes sign under the action of $g_1$ and $g_2$. 
  Since square or appropriate product of semi-invariants leads to invariant, we introduce  a new set of invariants:
\beqa
    a&=& w +  w' + \hat w +  \bar w \nn\\
   v&=&2 \mu + \lambda =
 ( w + \hat w)( w' +  \bar w)+2   w \hat w +2    w' \bar w\nn \\
   \tilde w^2&=&(w+ \hat w  - w' -\bar w)^2\nn\\
   j^2&=&( w - \hat w - w' + \bar w)^2( w -  \hat w +  w' - \bar w)^2\nn
 \eeqa
 which are respectively of degree $1,2,2,4$. Since the Jacobian
 \beqa
\frac{\partial(a,v,\tilde w^2, j^2)}{\partial(w,w',\bar w,\hat w)}=128 jj' \ ,\nn
\eeqa
the set of polynomial $(a,\tilde w^2,v,j^2)$ constitute an adapted set of primary invariants.
Now, the syzygy takes a more complicated form:
\beqa
{c}^{2}+ \left(\frac{{a}^{3}}8+\frac{a{{\it \tilde w}}^{2}} 8-\frac{av} 2 \right) c+{
\frac {{a}^{6}}{256}}+{\frac {{a}^{4}{{\it \tilde w}}^{2}}{128}}-\frac{{a}^{
4}v}{32}-\frac{{a}^{2}v{{\it \tilde w}}^{2}}{32}-{\frac {{j}^{2}{{\it \tilde w}}^{2}}{64}}+{
\frac {{a}^{2}{{\it \tilde w}}^{4}}{256}}+\frac{{a}^{2}{v}^{2}}{16}
=0 \ . \nn
\eeqa
This simply means that now any invariant polynomial writes
\beqa\label{eq:PI2}
I(w,w',\bar w,\hat w) = f(a,v,\tilde w^2, j^2) +  g(a,v,\tilde w^2, j^2) c \  .
\eeqa
This new set is more adapted for our purpose as we now show. Let
\beqa
\tilde d = \partial_{w} + \partial_{\hat w} -\partial_{w'}-\partial_{\bar w} \ . \nn
\eeqa
This operator has the property to map an invariant polynomial to a semi-invariant polynomial which picks up a sign  under $g_1$.  Moreover, it is immediate to observe that
\beqa
\tilde d(a)=\tilde d(j) = \tilde d(v) =0 \ \ \text{and} \ \ \tilde d (\tilde w) =4 \ . \nn
\eeqa
Observing that $\tilde d (c) =-2 g$, the method can be further improved eliminating $c$ in \eqref{eq:PI2} through the relation
\beqa
\label{eq:sub}
c=\frac{4a^2v+j^2 -a^4-16g^2} {16 a}\ , \ \ g=\frac{a \tilde w-j}4 \ . 
\eeqa
The substitution of \eqref{eq:sub} in \eqref{eq:PI2} gives for any invariant polynomial
\beqa
\label{eq:opt}
I(w,w',\bar w,\hat w) = f(a,v,\tilde w^2, j^2) +  g(a,v,\tilde w^2, j^2)  \frac{4a^2v+j^2 -a^4 -( a \tilde w - j)^2}{16a} = \frac {P(a,v,j,\tilde w)} a \ ,  
\eeqa
where now the polynomial $P$   is {\it uniquely defined} because $f$ and $g$ are unique. Moreover, $P$ depends on the invariant polynomials $(a,v)$ and on the semi-invariant polynomials $(j,\tilde w)$. The important property of our algorithm is that all these variables but $\tilde w$ are in the kernel of $\tilde d$. Since 
\beqa
\frac{\partial(a,v,j,\tilde w)}{\partial(w,w',\bar w,\hat w)}= -32(\bar w -w')(w-\hat w) \ , \nn
\eeqa
we are sure that the variables $(a,v,j,\tilde w)$ are algebraically independent.
We finally define 
\beqa
h=ac+g^2 \ , \nn
\eeqa
which is also in the kernel of $\tilde d$

We now apply the method to the second part of diagram $B$ which is the most difficult integral appearing in the whole three-loop calculation.
The weak field expansion of $I_B$ gives
    \beqa\label{eq:B}
    I_B^{\text{hard}}&=& -\rho \frac{\cosh\rho( w - w' + \hat w - \bar w)}{\cosh \rho w \cosh \rho w' \cosh \rho \bar w
      \cosh \rho \hat w }\Big[\frac 1 A -\frac C{G^2} \ln(1+\frac {G^2}{AC})\Big]\nn\\
    &=& \sum \limits_{n=0}^\infty \beta_n^{\text{hard}} \rho^{2n} \ .\nn
    \eeqa 
Now, $\beta_n$ develops using $1+\frac{g^2}{ac} = \frac 1 {1-\frac{g^2}h}$ as
       \beqa
    \beta_n^{\text{hard}}=\sum \limits_{k,\ell} u_n^{k,\ell}(w,w',\hat w, \bar w)
    \frac1 {g^k(h-g^2)^\ell} + \sum\limits_k
    v_n^k(w,w',\hat w, \bar w)
    \frac{\ln\Bigg(\frac 1 {1-\frac {g^2}{h}}\Bigg)}{g^k}\nn
    \eeqa
    Using \eqref{eq:opt}, we have
    \beqa
    u_n^{k,\ell}(w,w',\hat w, \bar w) &=& \frac {P_n^{k,\ell}(a,v,j,\tilde w)}a \ , \nn\\
    v_n^k(w,w',\hat w, \bar w)&=&\frac {P_n^{k}(a,v,j,\tilde w)}a  \nn \ , 
    \eeqa

    For the lowest order of the expansion we have
    \beqa
    \beta_0^{\rm hard} &=& u_0^{0,0} + v_0^0 \ln\Bigg(\frac 1 {1-\frac {g^2}{h}} \Bigg)+ v_0^2 \frac{\ln\Bigg(\frac 1 {1-\frac {g^2}{h}}\Bigg)}{g^2}  \ , \nn\\
    \beta^{\rm hard}_1
&=& u_1^{00} + u_1^{20} \frac 1 {g^2} + \ln\Big(1+ \frac {g^2}{ac}\Big)
\Bigg\{v_1^2 \frac 1 {g^2} + v_1^4 \frac 1 {g^4} \Bigg\} \ .\nonumber
\eeqa
with
\beqa
u_0^{0,0}= -\frac 1a \ , \ \
v_0^0 =-\frac 1a \ , \ \
v_0^2 = \frac h a \ , \nn
\eeqa
and
\beqa
u_1^{00}&=&\frac  a 6 -\frac {\tilde w^2} {2a} -\frac c {a^2} \ ,
\nonumber\\
u_1^{20}&=& -{\frac {{a}^{5}}{96}}-{\frac {{a}^{3}{{\it \tilde w}}^{2}}{192}}+\frac{{a}^
{3}v}    {16}-{\frac {{a}{{\it \tilde w}}^{4}}{192}}-\frac{{a}{v}^{2}}{12}+\frac{
{a}{{\it \tilde w}}^{2}v}{48}+{\frac {{a}{j}^{2}}{192}}+{\frac {{{\it \tilde w}}
^{2}{j}^{2}}{64a}}\nonumber\\
&&+c \left( -\frac{{a}^{2}}{16}-\frac{{{\it \tilde w}}^{2}}{6}+\frac{v}3 
 \right) \ , \nonumber\\
 v_1^2 &=&{\frac {5\,{a}^{5}}{768}}-{\frac {a{j}^{2}}{192}}-{\frac {{a}^{3}{{
\it \tilde w}}^{2}}{384}}-\frac{{a}^{3}v}{32}+{\frac {a{{\it \tilde w}}^{4}}{768}}+
\frac{a{v}^{2}}{48}+{\frac {a{{\it \tilde w}}^{2}v}{96}}+c \left( -{\frac {13\,{a}^{2
}}{24}}+{\frac {7\,{{\it \tilde w}}^{2}}{24}}+\frac{5v}6 \right) \ ,\nonumber\\
v_1^4&=&
{
\frac {7\,{a}^{7}{{\it \tilde w}}^{2}}{3072}}-{\frac {7\,{a}^{7}v}{768}}+{
\frac {5\,{a}^{5}{{\it \tilde w}}^{4}}{3072}}+{\frac {5\,{a}^{5}{v}^{2}}{192
}}+{\frac {{a}^{3}{{\it \tilde w}}^{6}}{3072}}-\frac{{a}^{3}{v}^{3}}{48}+{\frac {
av{j}^{2}{{\it \tilde w}}^{2}}{192}}-{\frac {{a}^{3}{{\it \tilde w}}^{4}v}{256}}-{
\frac {{a}^{3}{j}^{2}{{\it \tilde w}}^{2}}{256}}\nonumber\\
&&-{\frac {a{{\it \tilde w}}^{4}{j}^
{2}}{768}}+{\frac {{a}^{3}{{\it \tilde w}}^{2}{v}^{2}}{64}}-{\frac {5\,{a}^{
5}{{\it \tilde w}}^{2}v}{384}}+{\frac {{a}^{9}}{1024}}\nonumber\\
&&+c\left( {\frac {{a}^{6}}{64}}-\frac{{a}^{2}{{\it \tilde w}}^{2}v}{24}-\frac{{j}^{
2}v}{24}+\frac{{a}^{2}{v}^{2}}{12}+\frac{{a}^{2}{j}^{2}}{48}+\frac{{a}^{4}{{\it \tilde w}}^
{2}}{32}-
\frac{{a}^{4}v}{12}+{\frac {{a}^{2}{{\it \tilde w}}^{4}}{64}} \right) \ .\nonumber
\eeqa
It remains of course to make the substitution \eqref{eq:sub} for $c$.
Now using partial integration upon $\tilde d$ we can integrate out the variable $\tilde{w}$ and finally calculate $\beta_{0}^{\rm hard}$ and $\beta_1^{\text{hard}}$, see \cite{HRS}
\beqa
\int_{0}^{\infty}dw d w' d\hat w d\bar w \,\e^{-a} \beta^{\rm hard}_0 &=& 
-\frac{13}{12} + \frac{7}{8}\zeta_3
\nonumber\\
\int_0^{\infty} dw d w' d\hat w d\bar w \, \e^{-a} \beta_1^{\text{hard}} &=&  \phantom{-}\frac{121}{64} - \frac{203}{128}  \zeta_3 \ .
\nn
\eeqa

\section{Final Remarks}
The method presented here has proven effective and useful in the calculation of weak field coefficients of a Feynman diagram (diagram B) but is not necessarily restricted to such regime. Indeed, although we know not of a concrete example presently it is plausible that for certain class of Feynman diagrams their exact calculation might be carried out along very similar, if not almost identical, lines. This would imply that the scope of the method is farther reaching than has been exploited here. Nevertheless we must admit the speculative nature of this assertion and await for any such examples that may emerge in future investigations.


\section*{References}
\begin{thebibliography}{9}
 \bibitem{sturm-book}
Sturmfels B 2008 Algorithms in invariant theory 2nd edn. (Springer Berlin)

\bibitem{decker}
Decker W and  Lossen C 2006
Computing in algebraic geometry. A quick start using SINGULAR 
(Springer Berlin)

\bibitem{MSV}
Martin L,  Schubert C  and Villanueva Sandoval V M  2003
{\it Nucl. Phys. }{\bf B668}  335  hep-th/0301022

\bibitem{EH}
Heisenberg  W and  Euler H  1936 {\it Z. Phys.} {\bf 98}  714

\bibitem{Dittrich}
 Dittrich  W and  Gies  H  2000 
 {\it Springer Tracts Mod. Phys.} {\bf  166}  1
 
\bibitem{Dunne}
 Dunne G V  2002 {\it Cont. Adv. in QCD}  478  hep-th/0207046

\bibitem{GK}
 Gies H and   Karbstein F 2017  
{\it JHEP} {\bf  03}  108

\bibitem{DS}
 Dunne G V and    Schubert C 2000 
{\it Nucl.Phys.} {\bf  B564}  591

\bibitem{HMcS}
Huet I,  McKeon D G C  and  Schubert C 2010 
Quantum Field Theory Under The Influence Of External Conditions (Qfext09) (World Scientific) 

\bibitem{HMcS2}
  Huet I,    McKeon  D G G and  Schubert C 2010 {\it JHEP } {\bf 1012}  036 hep-th/1010.5315

\bibitem{AAM}
 Affleck I K,   Alvarez O and   Manton  N S 1982  {\it Nucl. Phys.} {\bf  B197}  509

\bibitem{Kras}
Krasnansky M  2008 {\it Int. J. Mod. Phys.} {\bf A23}   5201  hep-th/0607230

\bibitem{DunneSchubert}
  Dunne  G V  and Schubert C 2002  {\it Phys. Lett.}  {\bf B526} 55  hep-th/0111134

\bibitem{DS2}  
  Dunne  G V  and Schubert C  2002 {\it JHEP} {\bf 0208}  053  hep-th/0205004

 \bibitem{DS3}   
  Dunne  G V  and Schubert C  2002 {\it JHEP}  {\bf  0206}  042  hep-th/0205005


\bibitem{Cvitanovic}
Cvitanovic P   1977 {\it Nucl. Phys.} {\bf B127}   176

\bibitem{HRS}

 Huet I,  Rausch de Traubenberg M and   Schubert C  In preparation



\end{thebibliography}
\end{document}